*A new method using machine learning to integrate ECG and gated SPECT MPI for Cardiac Resynchronization Therapy Decision Support on behalf of the VISION-CRT*

**Authors**

Fernando de A. Fernandes[a], M.Sc.
fernando.fernandes2@gmail.com

Kristoffer Larsen[b],
kalarsen@mtu.edu

Zhuo He[c], BS
zhuoh@mtu.edu

Erivelton Nascimento[d], PhD., M.D.
hpcrates7@gmail.com

Amalia Peix[e], Ph.D., M.D.
atpeix@gmail.com

Qiuying Sha[b], PhD
qsha@mtu.edu

Diana Paez[f], MsEd., M.D.
d.paez@iaea.org

Ernest V. Garcia[g]. Ph.D.,

Weihua Zhou[c,h], PhD
whzhou@mtu.edu

Claudio T Mesquita[a], Ph.D., M.D.
claudiotinocomesquita@gmail.com

**Institutions**

a. Nuclear Medicine Department, Hospital Universitario Antonio Pedro-EBSERH-UFF, Niteroi, Brazil
b. Department of Mathematical Sciences, Michigan Technological University, Houghton, MI, USA
c. Department of Applied Computing, Michigan Technological University, Houghton, MI, USA
d. Cardiology Department, Hospital Universitario Antonio Pedro-EBSERH-UFF, Niteroi, Brazil
e. Nuclear Medicine Department, Institute of Cardiology, La Habana, Cuba
f. Nuclear Medicine and Diagnostic Imaging Section, Division of Human Health, Department of Nuclear Sciences and Applications, International Atomic Energy Agency, Vienna, Austria
g. Department of Radiology and Imaging Sciences, Emory University, Atlanta, GA

h.  Center for Biocomputing and Digital Health, Institute of Computing and Cybersystems, and Health Research Institute, Michigan Technological University, Houghton, MI, USA

Fernando de Amorim Fernandes and Kristoffer Larsen contributed equally to this work.

## *Funding*

Funding: This study presents the results derived from the International Atomic Energy Agency (IAEA) multicenter trial: ''Value of intraventricular synchronism assessment by gated-SPECT myocardial perfusion imaging in the management of heart failure patients submitted to cardiac resynchronization therapy'' (IAEA VISION-CRT), Coordinated Research Protocol E1.30.34, and received funds from IAEA. CTM receives grants from CNPq and FAPERJ. It was supported in part by a grant from The American Heart Association (Project Number: 17AIREA33700016, PI: Weihua Zhou) and by Michigan Technological University Undergraduate Research Internship Program (PI: Kristoffer Larsen).

### *Disclosure*
None of the authors have any conflicts of interest to disclose.

### *Running title*

*ML to predict CRT response*

### *Address for correspondence*

Fernando de Amorim Fernandes, Medical Physicist.
Tel.:  +55212629-9270 - +5521979820116
Email: fernando.fernandes2@gmail.com
Mailing address: 303 Marquês de Parana street - Niterói – Rio de Janeiro / Brazil - Postal code: 24033-900
Nuclear Medicine Department, Hospital Universitario Antonio Pedro-EBSERH-UFF

Or

Weihua Zhou
Tel: +1 906-487-2666
E-mail address: whzhou@mtu.edu
Mailing address: 1400 Townsend Dr, Houghton, MI 49931

**Background**

Cardiac resynchronization therapy (CRT) has been established as an important therapy for heart failure. Mechanical dyssynchrony has the potential to predict responders to CRT.

**Objectives**

The aim of this study was to report the development and the validation of machine learning (ML) models which integrates ECG, gated SPECT MPI (GMPS) and clinical variables to predict patients' response to CRT.

**Methods**

This analysis included 153 patients who met criteria for CRT from a prospective cohort study. The variables were used to modeling predictive methods for CRT. Patients were classified as ''responders'' for an increase of LVEF $\geq$ 5% at follow-up. In a second analysis, patients were classified ''super-responders'' for increase of LVEF $\geq$ 15%. For ML, variable selection was applied, and Prediction Analysis of Microarrays (PAM) approach was used for response modeling while Naïve Bayes (NB) was used for super-response. They were compared to models obtained with guideline variables.

**Results**

PAM had AUC of 0.80 against 0.71 of logistic regression with guideline variables (p = 0.47). The sensitivity (0.86) and specificity (0.75) were better than for guideline alone, sensitivity (0.72) and specificity (0.22). Neural network with guideline variables outperformed NB (AUC = 0.87 vs 0.86; p = 0.88). Its sensitivity and specificity (1.0 and 0.75, respectively) was better than guideline alone (0.40 and 0.06, respectively).

**Conclusions**

Compared to guideline criteria, ML methods trended towards improved CRT response and super-response prediction. GMPS had a central role in the acquisition of most parameters. Further studies are needed to validate the models.

*Abbreviations*

| | |
|---|---|
| *CABG* | *Coronary artery bypass graft* |
| *CAD* | *Coronary artery disease* |
| *CRT* | *Cardiac resynchronization therapy* |
| *ECTb4* | *Emory Cardiac Toolbox Version 4.0* |
| *ECG* | *Electrocardiogram* |
| *ESV* | *Left ventricular end systolic volume* |
| *GMPS* | *Gated myocardial perfusion SPECT* |
| *HF* | *Heart failure* |
| *IAEA* | *International Atomic Energy Agency* |
| *LBBB* | *Left bundle branch block* |
| *LV* | *Left ventricle* |
| *LVEF* | *Left ventricular ejection fraction* |
| *MI* | *Myocardial infarction* |
| *NYHA class* | *New York Heart Association Class* |
| *OSEM* | *Ordered subset expectation maximization* |
| *PCI* | *Percutaneous coronary intervention* |
| *PSD* | *Left ventricular phase histogram standard deviation* |
| *SPECT* | *Single photon emission tomography* |

## INTRODUCTION

Heart failure continues to be an increasingly prevalent disease with risk factors that vary substantially between geographies[1]. Cardiac resynchronization therapy (CRT) has been established as one of the most important therapies in the management of advanced symptomatic patient. For those with left bundle branch block (LBBB) and QRS duration > 150 ms, a strong recommendation for CRT implantation seems to be clear[2,3]. The reasons for the relevant percentage of patients that will not experience benefit from the procedure remains unclear.

Mechanical dyssynchrony as measured by phase analysis has emerged as a potential procedure to predict responders to CRT[4,5] and to improve lead positioning[6,7]. In addition, Gated myocardial perfusion SPECT (GMPS) imaging demonstrated value as an "all in one" tool for CRT[8,9]. Peix et al. have shown that changes in left ventricular phase standard deviation (PSD) before and after CRT were associated with response to the therapy[10]. However, two clinical trials were unable to demonstrate the benefits of early mechanical dyssynchrony to predict CRT response[10,11]. Ventricular remodeling was also supposed to be a potential marker of response to CRT, once its relation with cardiac function was well studied [12–14].

Considering the multifaceted and heterogeneous behavior of heart failure in combination with the capacity of machine learning algorithms to recognize patterns, it is hypothesized that a machine learning model could provide better prediction of CRT response. The aim of this study was to report the development and the validation of machine learning models integrates ECG, GMPS and clinical variables to predict patients' response to CRT.

## METHODS

*PATIENT POPULATION*

This is a post hoc analysis of a non-randomized, multinational, multicenter prospective cohort study: ''Value of intraventricular synchronism assessment by gated-SPECT myocardial perfusion imaging in the management of HF patients submitted to CRT'' (IAEA VISION-CRT) funded by the International Atomic Energy Agency (IAEA).

The inclusion criteria were symptomatic HF patients over 18 years old with NYHA functional class II, III or ambulatory IV HF for at least 3 months before enrollment despite optimal medical treatment according to the current guidelines; LVEF ≤ 35% from ischemic or non-ischemic causes measured according to the usual procedure at the participating center for inclusion, whereas LVEFs used for analysis came from the nuclear core lab; sinus rhythm with LBBB configuration defined as a wide QRSd (≥ 120 ms). Exclusion criteria were as follows: right bundle branch block, pregnancy or breast-feeding, acute coronary syndromes, coronary artery bypass grafting, or percutaneous coronary intervention in the last 3 months before enrolment and within 6 months of CRT implantation. The CRT devices were implanted using standard procedures. The LV lead was implanted in the posterolateral coronary vein, depending on vein availability.[10,15]

All patients provided written informed consent, patient anonymity was maintained during data analysis, and all procedures were done according to the Declaration of Helsinki.

Clinical characteristics and GMPS were assessed at baseline (before CRT) and at follow-up (6 ± 1 month after CRT). The patients were classified as ''responders'' or "non-responder" if they had an increase of LVEF ≥ 5% as measured by GMPS at follow-up. A second analysis was done by classifying patients as ''super-responders'' or "non-super-responders" to CRT if they had an increase of LVEF ≥ 15%.

*CLINICAL DATA*

A total of 12 variables were evaluated at baseline and at follow-up: patient characteristics as age, gender, and ethnicity; disease history as presence of CAD, previous myocardial infarction (MI), percutaneous coronary intervention (PCI), coronary artery bypass graft surgery (CABG), NYHA class, hypertension (HTM), diabetes (DM); smoking history; and medical treatment.

*SPECT MPI ASSESSMENT*

The GMPS was performed approximately 30 min post rest injection using 740 to 1110 MBq (20-30 mCi) of 99mTc-sestamibi or tetrofosmin. All the images were acquired on gamma cameras with high-resolution low energy collimators, using 180° orbits, 8 or 16 Image reconstruction was performed using the OSEM algorithm with three iterations and ten subsets, and a Butterworth filter with a power of 10 and a cutoff frequency of 0.3 cycles/mm. The resulting short-axis images were sent to Emory Cardiac Toolbox (ECTb4, Atlanta, GA) for automatized assessment of LV function including left ventricular ejection fraction (LVEF), left ventricular end-systolic volume (ESV), myocardial mass, stroke volume, wall thickening (WT), and scar; phase systolic and diastolic parameters including phase peak (PP), phase standard deviation (PSD), and phase bandwidth (PBW); and shape parameters such as end systolic eccentricity (ESE), end diastolic eccentricity (EDE), and shape index (ESSI and EDSI) analysis using the Emory Cardiac Toolbox (ECTb4, Atlanta, GA). Concordance between CRT LV lead position was recorded, and the recommended site was also considered. Concordance was defined as the agreement between CRT LV lead position recorded and the optimal LV lead position *identified by Emory Cardiac Toolbox* as the latest contracting viable site[11,16].

*MACHINE LEARNING*

A representation of the machine learning pipeline was presented in figure 1. It included data splitting (training and testing sets), feature selection, data imputation, synthetic minority oversampling technique (SMOTE), data transformations, model tuning within cross-validation, modeling, and prediction. The sample was split with 80% for training and 20% for testing. In figure 2, the flow chart displays the initial patients entering the study and the resulting samples used.

*VARIABLE SELECTION*

Five methods were used to select relevant variables to be used in the models. Information Gain, Recursive Feature Selection, Boruta, and Relief were performed individually. Afterwards, these feature subsets were culled using Pearson correlation (0.80) and near-zero variance filters. A fifth subset was created by simply taking the union of the four aforementioned feature selection subsets, and then again applying Pearson correlation and near-zero variance filters.

*MODEL BUILDING*

We used the Prediction Analysis of Microarrays (PAM)[17] aka "Nearest shrunken centroids (NSC)" to develop the predictive model that ultimately best fit CRT response using the selected variables. PAM creates centroids for each class, i.e., "response" and "non-response" in addition to an overall centroid encompassing all classes. The "threshold" is a tuning parameter that shrinks the feature centroids of each class toward the overall centroid for all classes in doing so reducing the effect of noise and removing variables which are unable to discriminate between classes. New samples are predicted by assigning the label of the nearest centroid using distance metrics, such as Euclidean distance.

For CRT super-response prediction, a Naïve Bayes (NB) model was utilized. It is a probabilistic classifier whose foundation is based on Bayes' theorem which also assumes a naïve assumption of feature independence. Using previously learned attributes, the model assigns class labels to problem instances using probability scores.

*CROSS VALIDATION*

Resampling was performed using 3-fold cross validation with 25 repeats. Within each iteration, SMOTE is applied to the training fold while data transformations are applied separately afterwards to both the training and testing fold. SMOTE is used to reduce model uncertainty and overly optimistic estimates of performance from overfitting, while also combating the issue of imbalanced classes. Data transformations include spatial sign, a technique of projecting features on a multidimensional sphere such that each sample is equidistant to the center of the distribution in order to reduce the effect of outliers, in addition to centering and scaling to pre-process the data.

*PREDICTION MODELS*

In order to compare the proposed models (PAM for response and NB for super-response to CRT), we trained two other models considering solely the variables proposed in the guideline: QRSd, LVEF, LBBB and NYHA[2]. Using the sample as described in the pipeline (figure 1), a logistic regression (LR-Guideline) was used for response prediction while a neural network for super-response (NNET-Guideline). The AUC, accuracy, sensitivity, and specificity from each model and from guideline recommendation criteria were compared.

*STATISTICAL ANALYSIS*

We compared the predictive performance of response and super-response for CRT of the ML models based on selected features with guideline recommendations and with models built over the variables proposed by the guideline[2]. Evaluation metrics used include area under the curve (AUC), accuracy, sensitivity, and specificity. For a direct comparison of the models, an AUC comparison test using Delong method was applied[18]. Statistical analysis was performed using R version 4.0.3 (2020-10-10).

*RESULTS*

Ten centers from 8 countries (Brazil, Chile, Colombia, Cuba, India, Mexico, Pakistan, and Spain) participated in IAEA VISION-CRT. A total of 199 patients were enrolled in the trial, 16 of them died before the follow-up or had greatly decreased ESV (ESV < 25 ml), which was an outlier caused by the low resolution of GMPS when measuring a small heart. Patients with missing data (LBBB, ECG QRS, NYHA, LVEF pre and post) were excluded from the Guideline models, but those with only missing pre and post LVEFs were considered for the ML models as presented in table 1.

The first endpoint of this study was response to CRT, defined as a difference greater than 5 percentage points (i.e. 5% increase) in LVEF from baseline to follow-up. The obtained percentage of responders (45.7%) was in agreement with the original results of Vision CRT clinical trial (47.0%)[10]. The second endpoint was super-response to CRT, defined as a difference greater than 15 percentage points (i.e. 15% increase) in LVEF. The actual super-responders (18.3%) in this study were also similar to the original results (21.5%)[19].

The variable importance obtained for each model was presented in figure 3 and 4, for PAM and NB models respectively.

The PAM model presented an AUC of 0.80 against 0.71 from the LR-Guideline model as shown in figure 5 (p = 0.47), all other parameters are presented in table 2. For PAM, the sensitivity (0.86) and specificity (0.75), were better than for guideline alone, sensitivity (0.72) and specificity (0.22).

When comparing super-response prediction models, the NNet-Guideline presented better performance in relation to NB model (AUC = 0.87 vs 0.86; p = 0.88) as shown in figure 6, all other parameters are presented in table 3. Considering the sensitivity and specificity, NNet-Guideline (1.0 and 0.75, respectively) was better than guideline alone (0.40 and 0.06, respectively).

## *DISCUSSION*

In this study we developed and validated a method for cardiac resynchronization therapy decision support which presented better response prediction performance in all metrics when compared to guideline recommendation. In a direct comparison, the PAM model presented accuracy of 0.80 vs 0.45 of the guideline, sensitivity of 0.86 vs 0.72, specificity of 0.75 vs 0.22. In AUC comparison, PAM model (0.80) was superior to logistic regression (0.71) using guideline parameters QRS, LBBB, LVEF, and NYHA, however the difference was not statistically significant (p = 0.47).

It is a machine learning model built on clinical variables and imaging data from both ECG and GMPS of 153 patients from the VISION clinical trial, a non-randomized, multinational, multicenter prospective cohort study. The clinical trial design included patients from different countries making the sample better represent the heterogeneity observed in practice[1]. The use of GMPS was also important to face the multivariable behavior of heart failure in a single and reproducible procedure[9]. A total of 20 features were extracted from SPECT images: 6 from LV function (LVEF, ESV, mass, stroke volume, and wall thickening), 10 from mechanical dyssynchrony (diastolic and systolic PP, PSD and PWB), and 4 from shape analysis.

Although early studies in favor of mechanical dyssynchrony and phase analysis have been not confirmed by clinical trials as independent predictors[5,10,20], our results showed their role in multivariable analysis. As presented in variable importance results (figures 3 and 4), Phase Peak and Phase Bandwidth had relevance and were included in the modeling process together with the usual variables.

In addition, the shape parameter EDE showed relevance. Recently, a post-hoc study has shown a potential applicability of shape parameters and ESE was also shown to be an independent predictor of CRT response in a multivariate analysis[21].

Another model was proposed for prediction of "super-response" to CRT. There is no clear definition for super-responders[22–25], however they can be understood as patients that present rare and remarkable improvements in the LV function after CRT and, consequently, reduced risk of subsequent cardiac events [22,25,26]. The option for LVEF alone as the marker of super-response was done considering that it's an objective and reproducible parameter[9] with LV functional information, feasible by echo, magnetic resonance, or GMPS. In our study, a LVEF increase of 15% was the criteria for super-response.

Only 27 of 150 patients (18%) presented super-response in our sample, similarly to other studies based on echo data that ranged from 16% to 25% of super-responders[22,25–27]. Guideline criteria for CRT seems to be not effective for the selection of super-responders; six patients did not have any guideline class, and, as demonstrated in table 3, and the accuracy (0.18), sensitivity (0.40) and specificity (0.06) were low. However, modeling a neural network with the same input variables (NNet-Guideline) resulted in substantial improvement for accuracy (0.79), sensitivity (1.00) and specificity (0.75). The Naïve Bayes model (NB) built from selected variables (AUC = 0.86) had no significant difference in comparison to NNet-Guideline (AUC = 0.87). We hypothesized that the differences between inclusion criteria (NYHA class, LVEF and QRS) and post-hoc results based on the same baseline parameters are mainly caused by the variabilities of different diagnostic methods and of different operators (intra and interobservers variation)[28].

It is well known that many complex problems do require complex solutions and it seems to be the case for heart failure patients' treatment. In the present work, we demonstrated that a multivariable approach can surpass single parameters or conventional criteria selection in the prediction of response and of super-response to CRT. Even with a relatively small and heterogeneous sample set, the machine learning models were able to combine clinical, functional, shape and phase data. In addition, Hung et al. (2021) demonstrated that LV dyssynchrony when evaluated in the viable myocardium, excluding scar, could be a better predictor that in the entire myocardium[29]. These promising early results suggests a need for continued research in these area as well as the controversial results about lead placement indication.

This study has several limitations. First of all, it was a post hoc analysis of a prospective non-randomized trial. Further prospective trials and external validations are expected. Another limitation was that all centers used the same software (ECTb) for cardiac orientation and parameters estimation. Because the approaches of other software providers to quantify LV function and dyssynchrony are different, the results from one cannot be directly translated to others[30,31]. Gated SPECT data was available to physicians before CRT implantation but it was not mandatory to follow these results guiding the CRT LV lead implantation. These may reduce the capacity to evaluate the clinical impact of the technique. The information provided by the short follow-up period also represents a limitation. Finally, the prognostic value of the models needs further investigation.

***CONCLUSIONS***

Compared to guideline criteria, ML methods trended towards improved CRT response and super-response prediction, combining clinical, functional, shape and phase data. GMPS had a central role in the acquisition of most parameters. Further studies are needed to validate the models.

***CLINICAL PERSPECTIVES***

CLINICAL COMPETENCIES.

Using multivariable data can be preferable to therapy decision of HF patients. LV functioning, shape and dyssynchrony as obtained by GMPS showed relevance for CRT when combined by machine learning tools.

TRANSLATIONAL OUTLOOK.

This work points in the direction of multivariable modeling for complex decision making in heterogenous populations. However future large prospective trials are needed to validate our findings. Additional dyssynchrony analysis on the viable myocardium portion can provide new relevant information.

*Acknowledgments*

CTM receives grants from CNPq and FAPERJ.

The authors thank the Vision CRT researchers for sharing the data and collaborate Amelia Jimenez-Heffernan, Sadaf Butt, Claudio T Mesquita, Teresa Massardo, Amalia Peix, Alka Kumar, Chetan Patel, Erick Alexanderson, Luz M Pabon, Ganesan Karthikeyan, Claudia Gutierrez, Ernest Garcia, and Diana Paez.

*Figure Titles and Legends*

*Tables (each on a separate page)*

Table 1. Baseline clinical characteristics

| Variables | Non-responder (n = 83, 54.3%) | Responder (n = 70, 45.7%) | Non-super-responder (n = 125, 81.7%) | Super-responder (n = 28, 18.3%) |
|---|---|---|---|---|
| Age, years | 59.8 ± 11.1 | 61.5 ± 10.2 | 60.9 ± 10.8 | 59.1 ± 10.4 |
| Females | 31 (37.3%) | 32 (45.7%) | 47 (37.6%) | 16 (57.1%) |
| Race | | | | |
| African | 9 (10.8%) | 6 (8.6%) | 14 (11.2%) | 1 (3.6%) |
| Asian | 5 (6.0%) | 1 (1.4%) | 5 (4.0%) | 3 (10.7%) |
| Caucasian | 11 (13.3%) | 10 (14.3%) | 18 (14.4%) | 3 (10.7%) |
| Hispanic | 47 (56.6%) | 34 (48.6%) | 68 (54.4%) | 13 (46.4%) |
| Indian | 11 (13.3%) | 19 (27.1%) | 20 (16.0%) | 10 (35.7%) |
| Smoking | 14 (16.9%) | 13 (18.6%) | 21 (16.8%) | 6 (21.4%) |
| DM | 24 (28.9%) | 14 (20.0%) | 36 (28.8%) | 2 (7.1%) |
| HTN | 53 (63.9%) | 38 (54.3%) | 79 (63.2%) | 12 (42.9%) |
| MI | 24 (28.9%) | 6 (8.6%) | 28 (22.4%) | 2 (7.1%) |
| CAD | 33 (39.8%) | 13 (18.6%) | 43 (34.4%) | 3 (10.7%) |
| CABG | 1 (1.2%) | 2 (2.9%) | 3 (2.4%) | 0 (0.0%) |
| NYHA | | | | |
| II | 17 (20.5%) | 27 (38.6%) | 30 (24.0%) | 14 (50.0%) |
| III | 58 (69.9%) | 34 (48.6%) | 79 (63.2%) | 13 (46.4%) |
| IV | 8 (9.6%) | 9 (12.9%) | 16 (12.8%) | 1 (3.6%) |
| ACEI/ARB | 62 (74.7%) | 63 (90.0%) | 101 (80.8%) | 24 (85.7%) |
| SPECT | | | | |
| ESV | 205.1 ± 103.3 | 171.9 ± 90.6 | 199.4 ± 102.9 | 147.4 ± 63.2 |
| LVEF | 28.3 ± 10.6 | 27.6 ± 11.9 | 28.1 ± 11.6 | 27.4 ± 9.2 |
| Mass | 223.8 ± 54 | 202.1 ± 50.2 | 219.6 ± 54.7 | 188.2 ± 40.3 |
| Stroke Volume | 70.5 ± 22.7 | 55.7 ± 18.5 | 66.8 ± 22.2 | 50.0 ± 13.4 |
| WT | 13.9 ± 8.5 | 13.2 ± 8.3 | 13.6 ± 8.6 | 13.7 ± 7.4 |
| Concordance | 20 (24.1%) | 16 (22.9%) | 29 (23.2%) | 7 (25.0%) |
| Scar | 27.9 ± 15.8 | 20.1 ± 11.9 | 25.3 ± 15.1 | 20.3 ± 11.6 |
| Diastolic | | | | |
| PBW | 171.3 ± 75.7 | 151.4 ± 80.2 | 168.6 ± 78.3 | 133.6 ± 72.2 |
| PK | 9.0 ± 8.3 | 9.1 ± 7.9 | 9.2 ± 8.6 | 8.4 ± 5.1 |
| PP | 226.2 ± 28.6 | 234.2 ± 32.7 | 226.3 ± 30.0 | 245.9 ± 28.4 |
| PSD | 52.7 ± 18.9 | 48.0 ± 22.0 | 52.2 ± 20.5 | 43.0 ± 18.8 |
| Systolic | | | | |
| PBW | 158.6 ± 73.8 | 139.6 ± 76.1 | 155.3 ± 76.2 | 126.1 ± 66.8 |
| PK | 8.6 ± 7.9 | 8.4 ± 8.0 | 8.8 ± 8.6 | 7.1 ± 3.0 |
| PP | 136.6 ± 26.2 | 145.8 ± 27.8 | 137.9 ± 26.9 | 153.9 ± 25.6 |
| PSD | 50.3 ± 19.9 | 46.0 ± 21.1 | 49.6 ± 21.2 | 42.7 ± 16.2 |
| EDE | 0.50 ± 0.20 | 0.55 ± 0.16 | 0.52 ± 0.18 | 0.56 ± 0.18 |
| EDSI | 0.84 ± 0.11 | 0.81 ± 0.09 | 0.83 ± 0.10 | 0.81 ± 0.09 |
| EDV | 275.5 ± 111.4 | 227.6 ± 96.0 | 266.2 ± 110.2 | 197.4 ± 68.6 |
| ESE | 0.54 ± 0.20 | 0.62 ± 0.13 | 0.57 ± 0.18 | 0.62 ± 0.12 |
| ESSI | 0.82 ± 0.12 | 0.77 ± 0.10 | 0.80 ± 0.12 | 0.77 ± 0.08 |
| ECG QRSd | 160.1 ± 27.5 | 158.9 ± 29.0 | 159.3 ± 30.3 | 160.8 ± 14.8 |

Values are n (%) or mean ± SD

DM = Diabetes mellitus; HTN = hypertension; MI = Myocardial infarction; CAD = Coronary artery disease; CABG = Coronary artery bypass graft; NYHA = New York Heart Association Functional Classification; ACEI = angiotensin-converting enzyme inhibitor; ARB = Angiotensin Receptor Blockers; SPECT = Single Photon Emission Computed Tomography; ESV = End-systolic volume; LVEF = Left Ventricular Ejection Fraction; WT = Wall thickening; Scar = % Non-viable LV; PBW = Phase bandwidth; PP = Phase peak; PK = Phase kurtosis; PSD = Phase Standard Deviation; EDE = End-diastolic eccentricity; EDSI = End-diastolic Shape Index; ESE = End-systolic eccentricity; ESSI = End-systolic shape index.

Table 2. Performance of response prediction models

| Model | Accuracy | Sensitivity | Specificity | AUC (95% IC) |
|---|---|---|---|---|
| Guideline | 0.45 | 0.72 | 0.22 | - |
| PAM | 0.80 | 0.86 | 0.75 | 0.80 (0.64 - 0.97) |
| LR-Guideline | 0.62 | 0.61 | 0.62 | 0.71 (0.51-0.91) |

PAM = Prediction Analysis of Microarrays; LR = Logistic Regression

Table 3. Performance of super-response prediction models

| Model | Accuracy | Sensitivity | Specificity | AUC (95% IC) |
|-------|----------|-------------|-------------|--------------|
| Guideline | 0.18 | 0.40 | 0.06 | - |
| NB | 0.80 | 0.80 | 0.80 | 0.86 (0.65-1.00) |
| NNet-Guideline | 0.79 | 1.00 | 0.75 | 0.87 (0.74-1.00) |

NB = Naïve Bayes; NNet = Neural Network

*Figures/Central Illustration*

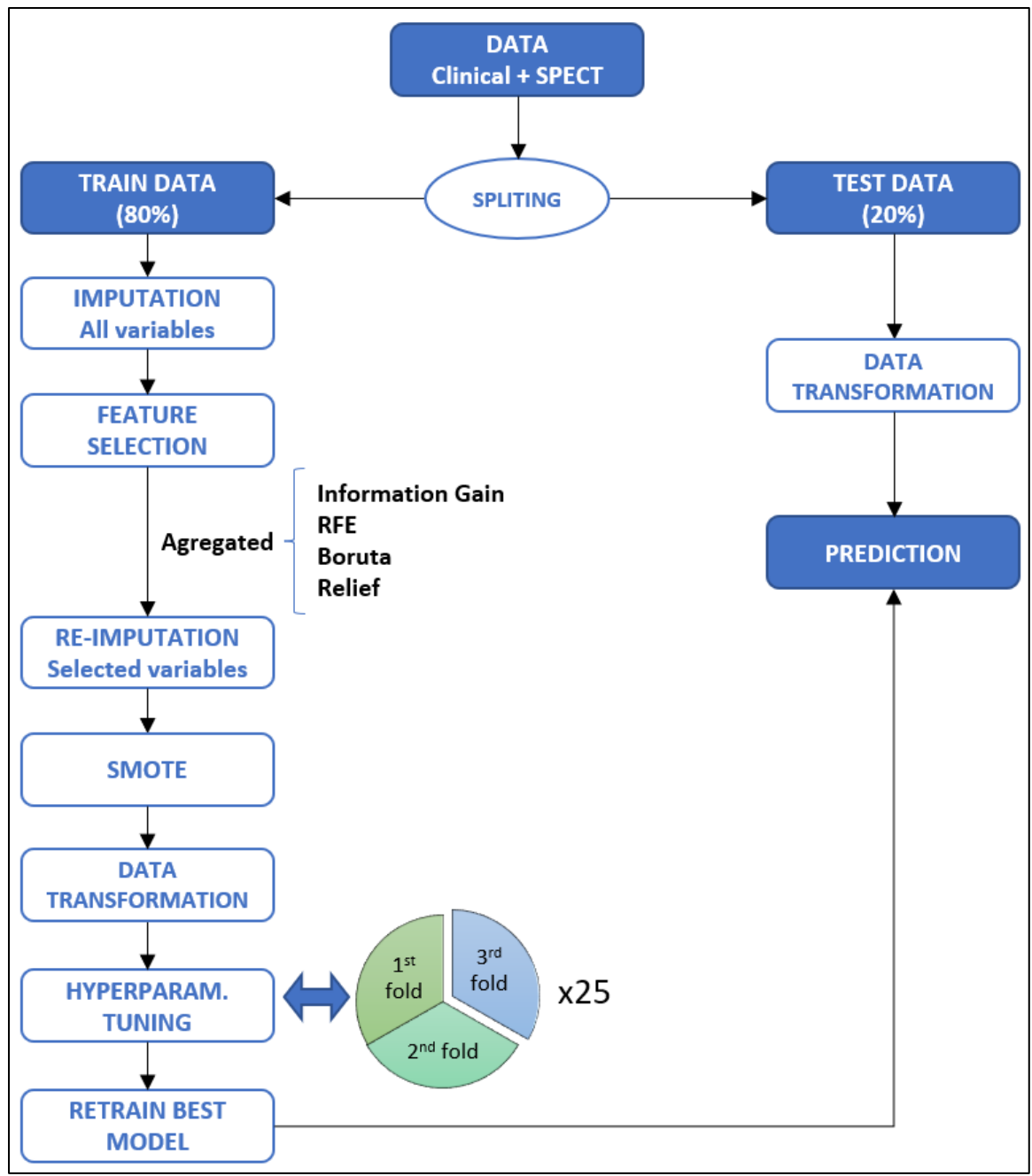

Figure 1. Modeling pipeline.

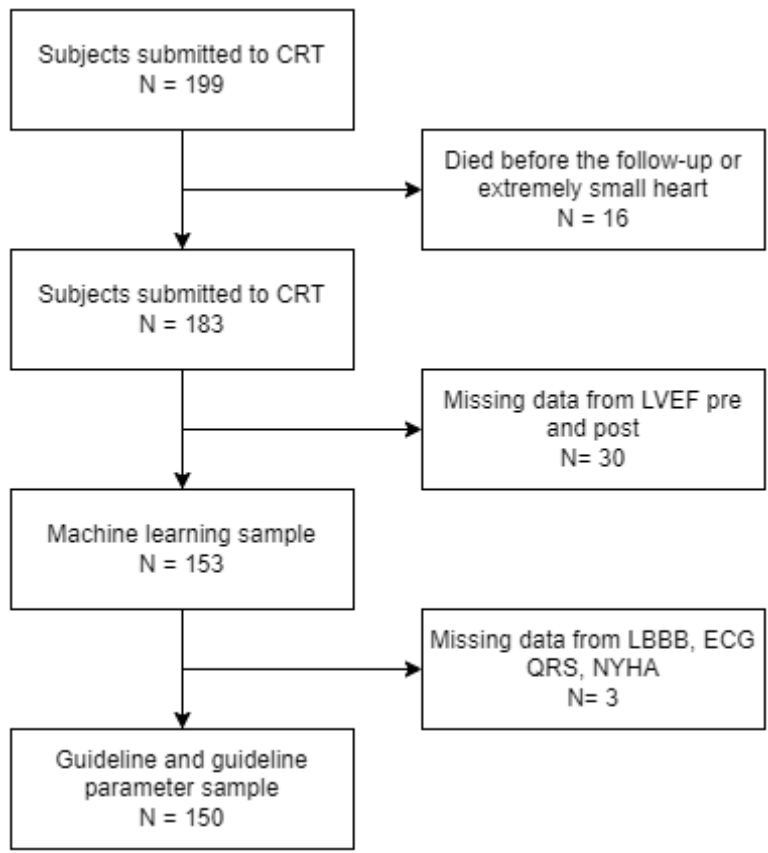

Figure 2. Study flow chart.

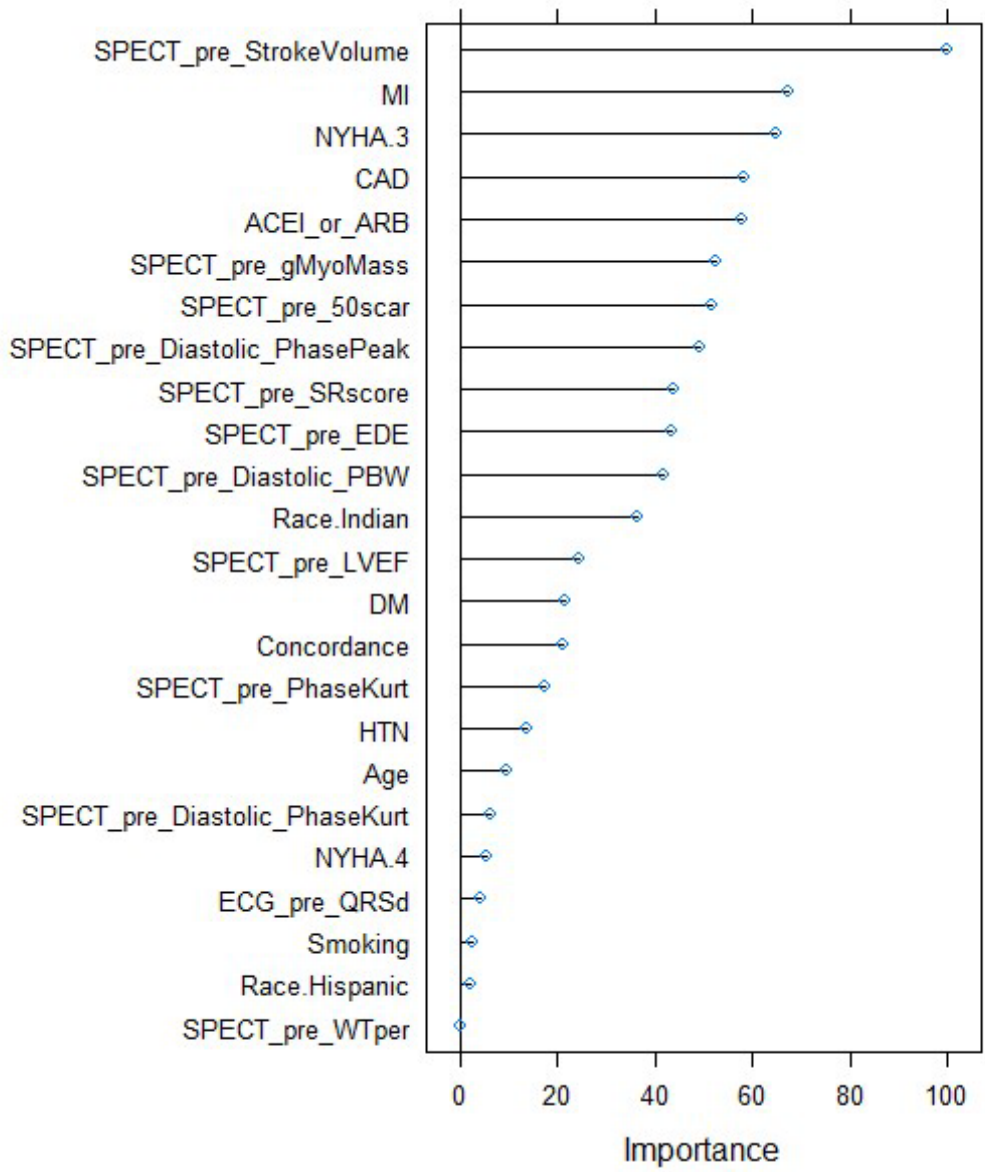

Figure 3. Importance of Variables for Prediction of CRT response.

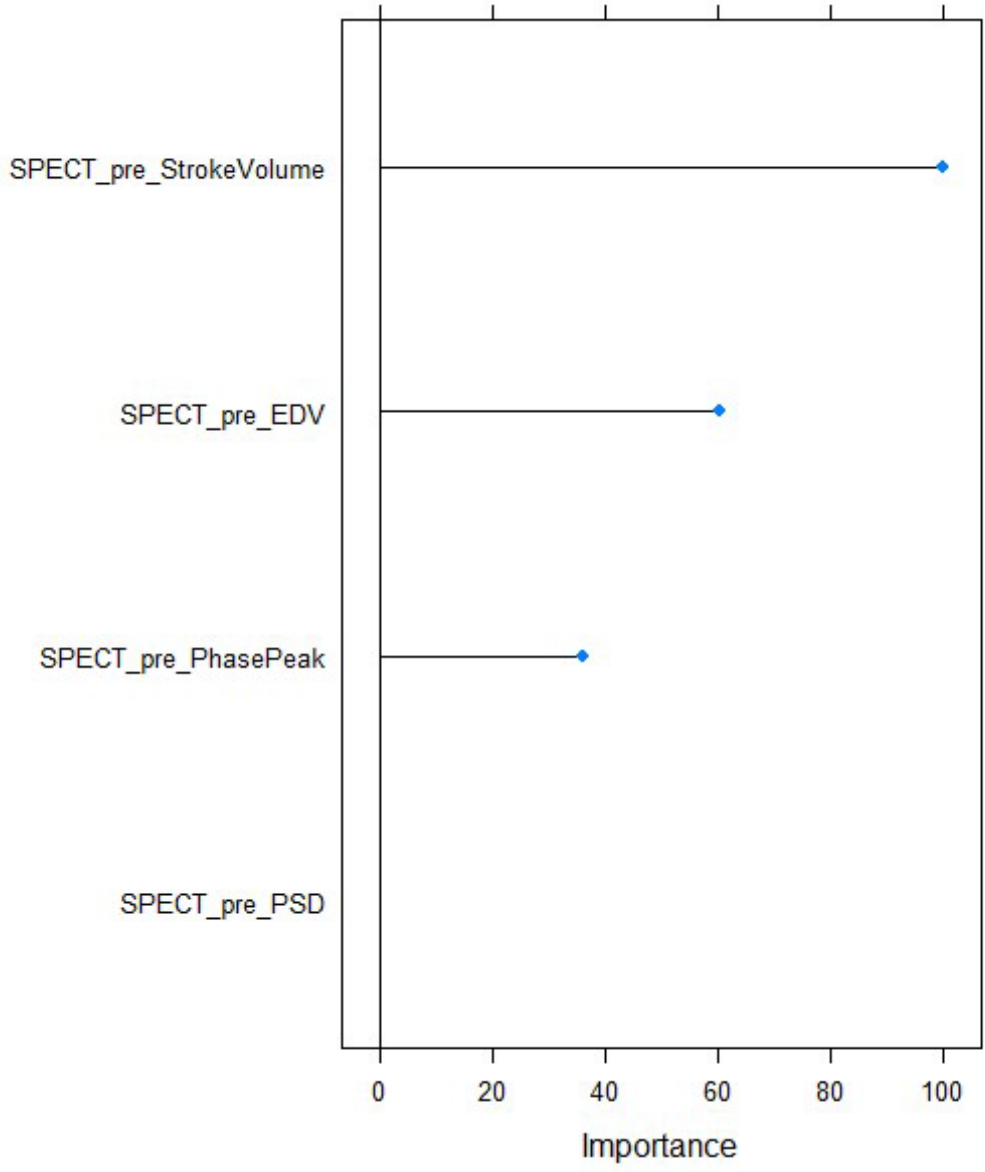

Figure 4. Importance of Variables for Prediction of CRT super-response.

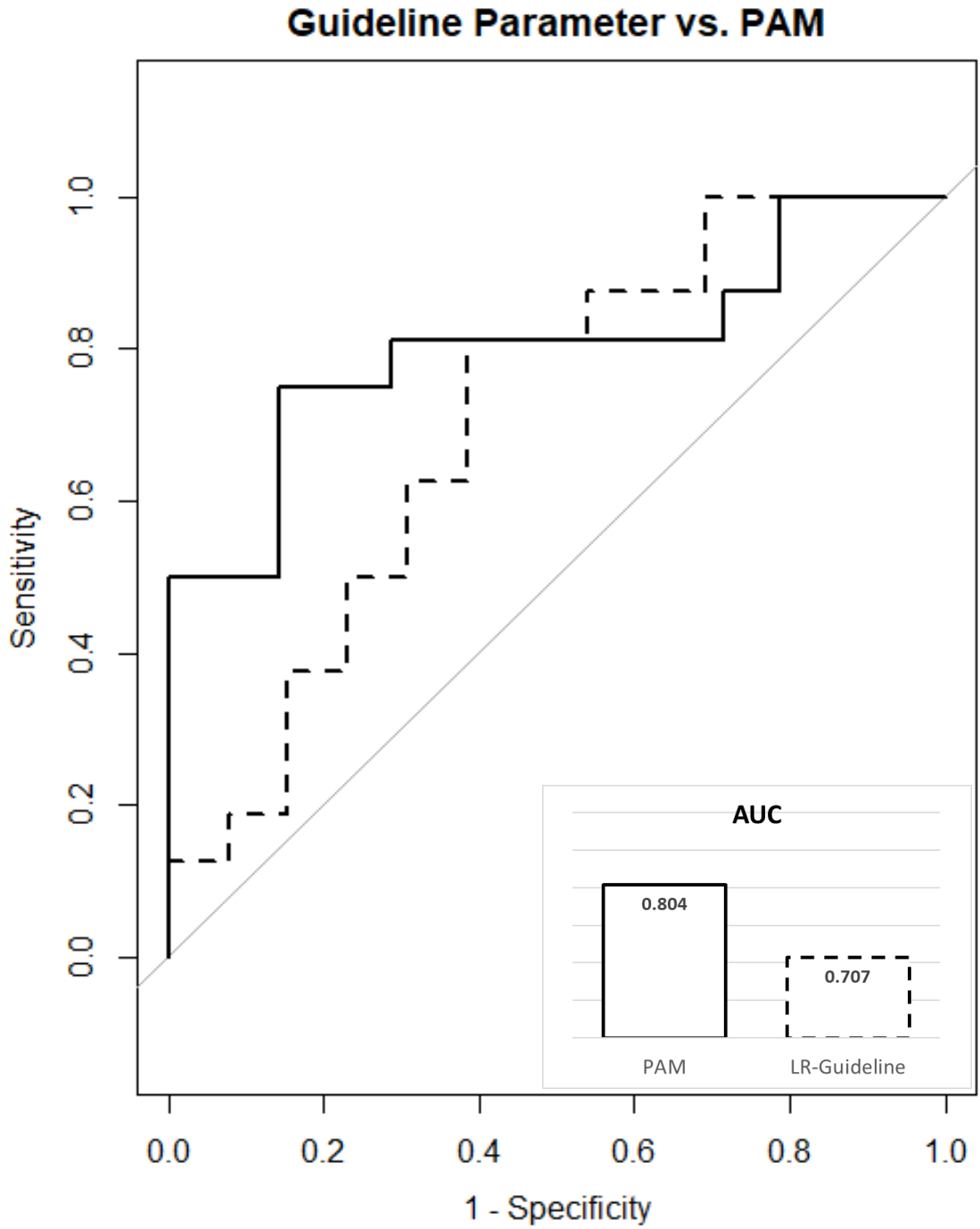

Figure 5. Receiver-Operating Characteristic Curves for Response to CRT

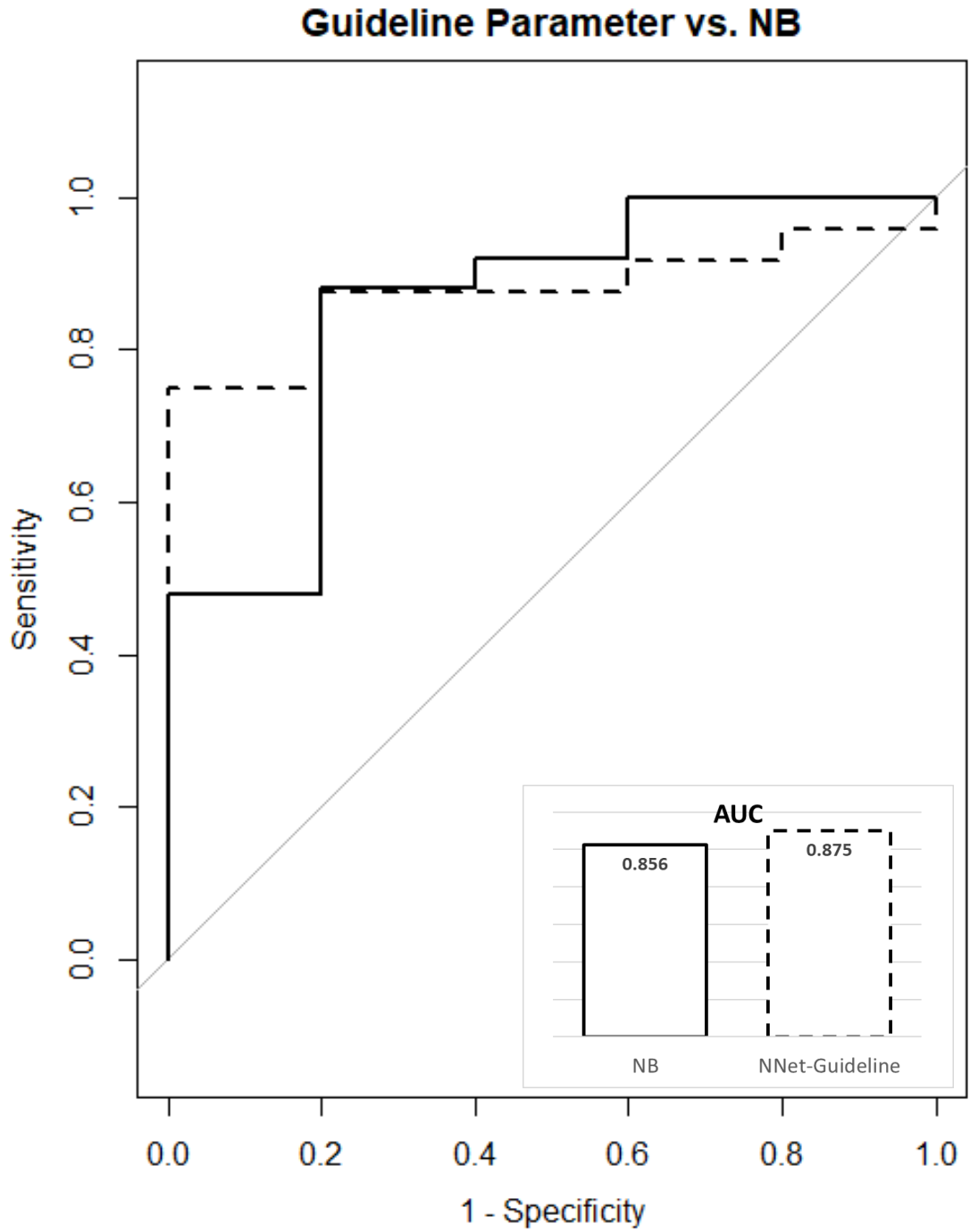

Figure 6. Receiver-Operating Characteristic Curves for Super-Response to CRT